\documentclass[12pt]{article}
\usepackage{graphicx}
\usepackage{graphics}
\usepackage{geometry}
\usepackage{float}
\usepackage{epsfig}
\usepackage{amssymb}
\usepackage{amsfonts}
\usepackage{amstext}
\usepackage{amsmath}
\usepackage{amsbsy}
\usepackage{latexsym}
\usepackage{dsfont}
\usepackage{fancyhdr}
\usepackage{colortbl}
\usepackage{psfrag}
\usepackage{subfigure}

\newwrite\ffile\global\newcount\figno \global\figno=1

\def\writedef#1{}

\input epsf
\def\figin{\epsfcheck\figin}\def\figins{\epsfcheck\figins}
\def\epsfcheck{\ifx\epsfbox\UnDeFiNeD
\message{(NO epsf.tex, FIGURES WILL BE IGNORED)}
\gdef\figin##1{\vskip2in}\gdef\figins##1{\hskip.5in}
\else\message{(FIGURES WILL BE INCLUDED)}%
\gdef\figin##1{##1}\gdef\figins##1{##1}\fi}

\def\figinsert{}
\def\ifig#1#2#3{\xdef#1{fig.~\the\figno}
\writedef{#1\leftbracket fig.\noexpand~\the\figno}%
\figinsert\figin{\centerline{#3}}\medskip\centerline{\vbox{\baselineskip12pt
\advance\hsize by -1truein\center\footnotesize{  Fig.~\the\figno.} #2}}
\bigskip\endinsert\global\advance\figno by1}
\def\endinsert{}

\begin{document}
\baselineskip 18pt
\newcommand{\Tr}{\mbox{Tr\,}}
\newcommand{\beq}{\begin{equation}}
\newcommand{\eeq}{\end{equation}}
\newcommand{\bdm}{\[}
\newcommand{\edm}{\]}
\newcommand{\bea}{\begin{eqnarray}}
\newcommand{\eea}[1]{\label{#1}\end{eqnarray}}
\renewcommand{\Re}{\mbox{Re}\,}
\renewcommand{\Im}{\mbox{Im}\,}

\def\N{{\cal N}}


\thispagestyle{empty}

\renewcommand{\thefootnote}{\fnsymbol{footnote}}

{\hfill \parbox{4cm}{
SHEP-05-11 \\
HU-EP-05-15 \\
MPP-2005-21
}}

\bigskip

\bigskip

\begin{center} \noindent \Large \bf

Strong coupling effective Higgs potential and a first order
thermal phase transition from AdS/CFT duality

\end{center}

\bigskip\bigskip\bigskip

\centerline{ \normalsize \bf Riccardo Apreda$^{a,b}$, Johanna Erdmenger$^{a,b}$,
Nick Evans$^c$ and Zachary Guralnik$^b$
 \footnote[1]{\noindent \tt apreda@df.unipi.it,
jke@mppmu.mpg.de,
evans@phys.soton.ac.uk, 
 zack@physik.hu-berlin.de } }

\bigskip

\bigskip\bigskip

\centerline{ $^a$ {\it Max Planck-Institut f\"ur Physik
(Werner Heisenberg-Institut)} }
\centerline{\it  F\"ohringer Ring 6, D - 80805 M\"unchen, Germany}

\bigskip

\bigskip

\centerline{ $^b$ {\it Institut f\"ur Physik, Humboldt-Universit\"at zu Berlin} }
\centerline{\it  Newtonstra\ss e 15, D - 12489 Berlin, Germany}

\bigskip

\bigskip

\centerline{ $^c$ {\it Department of Physics, Southampton University} }
\centerline{\it  Southampton  SO17 1BJ, United Kingdom}
\bigskip

\bigskip\bigskip

\bigskip

\renewcommand{\thefootnote}{\arabic{footnote}}

\centerline{\bf \small Abstract}

\medskip

{\small \noindent  }
We use AdS/CFT duality to study the
 thermodynamics of a strongly coupled ${\cal N} =2$
supersymmetric large $N_c$ $SU(N_c)$ gauge theory with $N_f =2$
fundamental hypermultiplets.  At finite temperature $T$ and
isospin chemical potential $\mu$, a potential on the Higgs branch
is generated, corresponding to a potential on the moduli space of
instantons in the AdS description. For $\mu =0$, there is a known
first order phase transition around a critical temperature $T_c$.
We find that the Higgs VEV is a suitable order parameter for this
transition; for $T>T_c$, the theory is driven to a non-trivial
point on the Higgs branch. For $\mu \ne 0$ and $T=0$, the Higgs
potential is unbounded from below, leading to an instability of
the field theory due to Bose-Einstein condensation.

\bigskip \bigskip
\bigskip\bigskip
\bigskip\bigskip 

\newpage


\section{Introduction}

In its original form, AdS/CFT duality
\cite{Maldacena:1997re,Gubser:1998bc,Witten:1998qj} relates
theories of closed strings in asymptotically AdS spaces to large
$N_c$ gauge theories with matter in the adjoint representation.
Fields in the fundamental representation may be added by including
an open string sector through the introduction of branes probing
the supergravity background. Much effort has gone into studying
dualities of this type, motivated largely by the goal of finding a
supergravity background dual to QCD.

The first example of AdS/CFT duality for a theory with
fundamental
representations related a conformal ${\cal N}=2$ $Sp(N)$
gauge
theory to string theory in $AdS_5 \times S^5/Z_2$, with
D7-branes
wrapping the $Z_2$ fixed surface with geometry $AdS_5
\times S^3$
\cite{Fayyazuddin:1998fb,Aharony:1998xz}. A different approach to flavour in AdS/CFT has been considered in \cite{Bertolini:2001qa}. In
\cite{Karch:2002sh},
the duality of \cite{Fayyazuddin:1998fb,Aharony:1998xz} was extended to an ${\cal N}=2$ $SU(N_c)$
theory with
$N_f$ massive fundamental hypermultiplets, essentially by
removing
the $\mathbb{Z}_2$ orientifold, which was justified by the fact that
a
probe D7-brane wrapping a contractible $S^3$ does not lead
to a
tadpole requiring cancellation.  Although the field theory
is not
asymptotically free, it has a UV fixed point in the strict
$N_c
\rightarrow\infty$ limit.   Following this, there have been a
number
of papers generalizing the duality to confining theories
with
fundamental representations
\mbox{\cite{Kruczenski:2003be}
-\cite{
Siopsis:2005yj},} including non-supersymmetric examples in which
spontaneous chiral symmetry breaking by a $\bar\psi\psi$ quark
condensate occurs \cite{BEEGK} -\cite{
Evans:2005ti}.

At zero temperature and vanishing quark mass, the ${\cal N}=2$
gauge theory has a non-trivial Higgs branch, i.e. a moduli space
of vacua on which the scalar components of the fundamental
hypermultiplets have expectation values.  For finite quark mass,
the moduli space includes a mixed Coulomb-Higgs branch: the
fundamental hypermultiplets may have non-zero expectation values
if the vector multiplet scalars have particular VEV's equal to the
quark mass. For both the massless and massive case, the moduli
space of field theory is described in the AdS picture by instanton
configurations on the D7-branes
\cite{Guralnik:2004ve,Guralnik:2004wq,Guralnik:2005jg,Erdmenger:2005bj}.
Self dual field strengths are solutions of the D7-brane equations
of motion due to a conspiracy between the Yang-Mills and
Wess-Zumino terms in the D7-brane action.  In general, this
conspiracy is destroyed by supersymmetry breaking deformations, which
give rise to a potential on the moduli space of instantons,
corresponding to an effective potential on the Higgs branch. We
will compute this potential at finite isospin chemical potential
and temperature. We  focus on the potential generated on a slice
of the moduli space corresponding to a single instanton centered
at the origin. This slice is parameterized by the instanton size,
which is dual to a particular Higgs VEV.

At finite isospin chemical potential,  we find the expected result
that the theory is destabilized via an effective negative mass
squared for the moduli, leading to Bose-Einstein condensation. The
negative mass squared term in the effective potential is related
to the metric on the Higgs branch, which is correctly reproduced
by the dynamics of spinning instantons on the D7-branes.  To
stabilize the theory, a positive scalar mass which is larger than
the chemical potential would need to be introduced.

Next we
consider finite temperature deformations.
Some finite temperature properties of theories with fundamental
representations have been studied previously using AdS/CFT duality
\cite{BEEGK,Kruczenski:2003uq,Ghoroku:2005tf}. As shown in
\cite{BEEGK,Kirsch:2004km}, there is a finite temperature first
order phase transition as the ratio of the temperature to the
quark mass is varied in the ${\cal N}=2$ theory of
\cite{Karch:2002sh} (a similar transition was discussed in
\cite{Kruczenski:2003uq}). In the AdS description, this transition
corresponds to a change in the topology of the D7-brane embedding
in an AdS-Schwarzschild background.

We explore the finite temperature behavior of this theory in more
detail by computing the effective potential generated on the Higgs
component of the moduli space. We find that the Higgs expectation
value is an order parameter for the first order phase transition.
At $0<T \le T_c$ and $\mu =0$, the Higgs VEV is driven to the
origin of moduli space. However for $T>T_c$ we find a surprise;
the instanton size is driven towards a non-zero value, suggesting
the existence of a vacuum in which the theory is higgsed. This is
in contrast to the weak coupling behavior of the theory,  for
which the one-loop finite temperature effective potential implies
that the origin of moduli space is at least metastable.

The organisation of this paper is as follows.  In section 2, we
review the AdS description \cite{Guralnik:2004ve,Guralnik:2004wq}
of the mixed Coulomb-Higgs branch in the Karch-Katz ${\cal N}=2$
theory \cite{Karch:2002sh}. In section 3, we consider the same
theory at finite chemical potential. Section 4 is devolved to the
finite temperature case.  In section 5 we briefly summarize our
results and discuss future developments.

\section{SUGRA dual of an ${\cal N}=2$ theory with fundamental representations}

We consider an ${\cal N}=2$ gauge theory which is dual
\cite{Karch:2002sh} to string theory in $AdS_5 \times S^5$ with
$N_f$ D7-branes wrapping a surface which is asymptotically $AdS_5
\times S^3$.  The matter content of this gauge theory is that of
the ${\cal N}=4$ $SU(N_c)$ gauge theory together with $N_f$
massive hypermultiplets in the fundamental representation. In
${\cal N}=1$ superspace, the Lagrangian is
\begin{align}{\cal L} = &{\rm Im}\left[\tau \int d^2 \theta d^2 \bar\theta
\left({\rm tr}\, (\bar \Phi_I e^V \Phi_I e^{-V}) + Q_i^\dagger
e^V Q^i + \tilde Q_i^\dagger e^{-V} \tilde Q^{i} \right) \right. \nonumber \\
&+ \left. \tau \int d^2 \theta \left({\rm tr}\, ({\cal W}^\alpha
{\cal W}_\alpha) + W\right) + \tau \int d^2\bar{\theta} \left({\rm
tr}\, ( \overline{{\cal W}}_{\dot{\alpha}} \overline{ {\cal
W}}^{\dot{\alpha}} ) +{ \overline{W}}\right) \right] ,
\end{align}
where the superpotential $W$ is
\begin{align}
W= {\rm tr} (\epsilon_{IJK}\Phi_I\Phi_J\Phi_K) + \tilde Q_i (m +
\Phi_3) Q^i .
\end{align}
The superfields $Q^i$ and $\tilde Q_i$, labeled by the flavor
index  $i=1 \cdots N_f$, comprise the ${\cal N}=2$ fundamental
hypermultiplets.

This theory is not asymptotically free and, at finite $N_c$, the
corresponding string background suffers from an uncancelled
tadpole.  However, as in \cite{Karch:2002sh}, we focus strictly on
the $N_c\rightarrow\infty$ limit with fixed $N_f$. In this case
there is a non-trivial UV fixed point for the 't Hooft coupling,
while the dual AdS string
background does not suffer from a tadpole problem since the probe
D7-branes wrap a contractible $S^3$.

In coordinates which will be convenient for our purposes, the
$AdS_5 \times S^5$ background is
\begin{align} \label{adsgeo}
ds^2 &= \frac{r^2}{R^2} dx^\mu dx^\mu + \frac{R^2}{r^2} (
d y^2 + y^2 d\Omega_3^2 + \sum_{i=1}^2 dZ^idZ^i),\\
e^{\Phi} &=g_s, \nonumber \\
F_{(5)}& = dC_{(4)} = 4 R^4 ( V_{S^5} + ^*V_{S^5}),
\qquad
C_{(4)}|_{0123} = {r^4 \over R^4 }\, dx^0\wedge dx^1 \wedge dx^2 \wedge dx^3, \nonumber\\[1mm]
\qquad R^4 &= 4 \pi g_s N_c \alpha^{'2},  
\nonumber
\end{align}
where $V_{S^5}$ is the volume form on $S^5$, and $r^2\equiv y^2 + Z^iZ^i$. 
The 't Hooft coupling in the dual gauge theory is $\lambda =
g^2N_c = g_sN_c$.

\definecolor{headercolor}{rgb}{0.65,0.65,0.9}
\definecolor{bgcolor}{rgb}{0.9,0.9,1.0}
\definecolor{cellcolor}{rgb}{0.75,0.75,0.85}
\newcolumntype{C}{>{\columncolor{cellcolor}}c}
\begin{table}
\begin{center}
\renewcommand{\arraystretch}{1.2}
\setlength{\arrayrulewidth}{.3\tabcolsep} \arrayrulecolor{bgcolor}
\begin{tabular}{*{10}{>{\columncolor{bgcolor}}c}}
        \multicolumn{10}{>{\columncolor{headercolor}}c}{Coordinates}\\
        \rowcolor{white}
        0&1&2&3&4&5&6&7&8&9\\
        \multicolumn{4}{C}{D3}&&&&&&\\ \hline
        \multicolumn{8}{C|}{D7}&&\\ \hline
        \multicolumn{4}{C}{$x^\mu$}&
                \multicolumn{4}{|C|}{$y^m$ (or $y,\Omega_3$) } &
                \multicolumn{2}{C}{$Z^i$}\\ \hline
        &&&&\multicolumn{6}{|C}{$r$}\\ \hline
\end{tabular}
\arrayrulecolor{black} \caption{\label{tab:indices}Index
conventions}
\end{center}
\end{table}

The fundamental hypermultiplets arise from $N_f$  D7-branes
embedded in this geometry. Their action is
\begin{equation}
S_{D7} ~ = ~ -T_7 \int d^4x d^4y \sqrt{- \det G_{ab}|_{PB}} ~ = ~
-T_7 \int ~d^4x ~dy ~d \Omega_3 ~y^3 \sqrt{1 + (Z_i^{'})^2},
\end{equation}
where $G_{PB}$ is the pull-back of the metric, a prime indicates a
derivative with respect to $y$, and we have assumed an embedding
independent of the coordinates $x$ as well as the coordinates on
$S^3$. There is a manifest minimum of the action when $Z'_i=0$. We
will choose
\begin{align}\label{emb}Z^1 =m, \quad Z^2=0\, ,
\end{align}
other solutions being related by the $U(1)_R$ symmetry
corresponding to rotations in the plane spanned by the $Z^i$. The
induced metric is
\begin{align}\label{induc} ds^2 = {r^2\over R^2} dx_{//}^2 &+ {R^2
\over r^2} ( d y^2 + y^2 d\Omega_3^2),
 \end{align} with $r^2 = y^2 + m^2$. The
parameter $m$ corresponds to the mass of the fundamental
hypermultiplets. For $m=0$, the geometry is $AdS_5 \times S^3$,
while for $m\neq 0$, the geometry approaches $AdS^5 \times S^3$ at
large $r$.  The $S^3$ component of the D7-geometry contracts to
zero size at $r=m$. So long as $N_f$ is held fixed in the limit
$N_c\rightarrow\infty$ with fixed $\lambda = g_s N_c \gg 1$, one
can neglect the back-reaction of the D7-branes on the bulk
geometry.

\subsection{The Higgs branch}

When the theory has massless quarks, the fundamental scalars $q^i$
and $\tilde q_i$ (denoting bottom components of chiral superfields
by lowercase letters,) have non-zero expectation values on the
Higgs branch\footnote{The scalars $\phi_1$ and $\phi_2$ belonging
to the adjoint hypermultiplet may also have non-zero expectation
values on the Higgs branch.} while the adjoint scalar $\phi_3$ of
the ${\cal N}=2$ vector multiplet vanishes. For non-zero mass,
$m$, and vanishing $\phi_3$,  the fundamental hypermultiplets are
massive and a pure Higgs branch does not exist.  However there is
a mixed Coulomb-Higgs branch when $\phi_3$ has an expectation
value such that some components of the hypermultiplets are
massless. An example of a point on a mixed Coulomb-Higgs branch is
given by a diagonal $\phi_3$ for which all but the last $k$
entries are vanishing,
\begin{align}\label{Coul}
  \phi_3 = \begin{pmatrix} 0& & & & & \cr &\ddots & & & &  \cr
        & &0 & & & \cr & & &-m & &  \cr & & & &\ddots & \cr
        & & & & & -m
\end{pmatrix}\, .
\\[1cm]\nonumber
\end{align}
In this case, the F-flatness equations $\tilde q_i (\phi_3+m) =
(\phi_3+m) q^i = 0$ permit fundamental hypermultiplet expectation
values in which only the last $k$ entries of $q^i$ and $\tilde
q_i$ are non-zero,
\begin{align}\label{cond}q^i= \begin{pmatrix} 0 \cr \vdots \cr 0 \cr \alpha_1^i \cr
\vdots \cr \alpha_k^i \end{pmatrix}\, , \qquad  \tilde q_i =
\begin{pmatrix} 0 & \cdots & 0 & \beta_{1i} & \cdots
& \beta_{ki} \end{pmatrix}\, .\end{align}  There are additional F
and D-flatness constraints  which we have not explicitly written.

In string theory, nonzero entries in (\ref{cond}) physically
correspond to D3-branes which are coincident with and dissolved
within the D7-branes. Dissolved D3-branes can be viewed as
instantons in the eight-dimensional world-volume theory on the
D7-branes \cite{Douglas:1995bn}, due to the Wess-Zumino coupling
\begin{equation}
S_{WZ} = \frac{T_7}{g_s} (2\pi\alpha')^2 \int C|_{PB}^{(4)}
   \wedge {\rm tr} (F\wedge F) \, .
\end{equation}
There is a one to one map between the moduli space of Yang-Mills
instantons and the Higgs branch of this ${\cal N}=2$ theory.   The
ADHM constraints from which instantons are constructed \cite{ADHM}
are equivalent to the F and D-flatness equations
\cite{Witten:1995gx,Douglas:1996uz}
 (see also \cite{Dorey:2002ik} for a review).

\subsection{Supergravity description of the Higgs branch}
\label{sughiggs}

 Because of the known one-one correspondence
between instantons and the Higgs branch,  one expects that
instantons solve the equations of motion of the non-Abelian\footnote{The existence of a Higgs branch requires at least two
flavors, or two D7-branes.} Dirac-Born-Infeld action describing
D7-branes embedded in (\ref{adsgeo}) according to (\ref{emb}). The
existence such solutions is a non-trivial consequence of AdS/CFT
duality \cite{Guralnik:2004ve,Guralnik:2004wq,proceedings}.

The effective action describing D7-branes in the AdS background
\eqref{adsgeo} is
\begin{align}\label{ac}
S &= - \, T_7 (2\pi\alpha')^2 \left( - \frac{1}{g_s} \int\,d^{8}\xi\,  C^{(4)} \wedge{\rm tr}(F
\wedge F)  + \int\,d^{8}\xi\, \frac{ e^{-\Phi}}{4}\sqrt{-\det G} \, \,{\rm
Tr}\left( F_{\alpha\beta}F^{\alpha\beta}
\right)\right) + \cdots \, , 
\end{align} where we have not written terms involving fermions and scalars.
This action is the sum of a Wess-Zumino term, a Yang-Mills term,
and an infinite number of corrections at higher orders in
$\alpha'$ indicated by $\cdots$ in \eqref{ac}.  The correspondence
between instantons and the Higgs branch suggests that the
equations of motion should be solved by field strengths which are
self-dual with respect to a flat four-dimensional metric.
\newpage
In this
paper, we work to leading order only in the large 't Hooft
coupling expansion generated by AdS/CFT duality, which allows us
to neglect
the higher order terms in the $\alpha'$
expansion\footnote{The dimensionless expansion parameter is
$\alpha'/R^2 = 1/\sqrt{\lambda}$.} of the action. Constraints on
unknown higher order terms arising from the existence of instanton
solutions, as well as the exactly known metric on the Higgs
branch, were discussed in \cite{Guralnik:2004ve,proceedings}.

The induced metric \eqref{induc} can be written as
\begin{align} ds^2 = {r^2\over R^2} dx^\mu dx^\mu + {R^2 \over r^2}
\sum_{m=1}^4 dy^m dy^m \, ,
\end{align}
with $r^2 = y^m y^m + m^2$.   Field strengths which are self dual
with respect to the flat four-dimensional metric $ds^2 =
\sum_{m=1}^4 dy^mdy^m$ solve the equations of motion, due to a
conspiracy between the Wess-Zumino and Yang-Mills term. Inserting
the explicit AdS background values \eqref{adsgeo} for the metric
and Ramond-Ramond four-form,
into the action \eqref{ac} for D7-branes embedded according to
\eqref{emb}, with non-trivial field strengths only in the $y^m$
directions, gives
\begin{align}\label{theactn}
\begin{split}
  S &= -T_7 (2\pi\alpha')^2 \int\, d^4x\,d^4y\, \frac{r^4}{4 g_s R^4} \,
       \left(-\frac{1}{2}\epsilon_{mnrs}F_{mn}F_{rs} +
       F_{mn}F_{mn}\right) \\
    &= -T_7 (4\pi\alpha')^2 \int \, d^4x\,d^4y\,  \frac{r^4}{4 g_s R^4} F_-^2 \, ,
\end{split}
\end{align}
where 
$F^-_{mn} = \frac{1}{2}(F_{mn}-\frac{1}{2}\epsilon_{mnrs}F_{rs})$.
Field strengths $F^-_{mn} = 0$, which are self-dual with respect
to the flat metric $dy^mdy^m$, manifestly solve the equations of
motion\footnote{Anti-instantons, with $F^+ =0$, correspond to
non-supersymmetric configurations which do not solve the equations
of motion.}. These solutions correspond to points on the Higgs
branch of the dual ${\cal N}=2$ theory. If $m\ne 0$ this is,
strictly speaking, a point on the mixed Coulomb-Higgs branch, with
expectation values of the form \eqref{Coul}, \eqref{cond}. We
emphasize that in order to neglect the back-reaction due to
dissolved D3-branes, we are considering a portion of the moduli
space for which the instanton number $k$ is fixed in the large
$N_c$ limit.

\subsection{A slice of the Higgs branch}

For simplicity, we consider the case $N_f=2$, which is the minimum
value for which a non-trivial Higgs branch exists.  We will focus
on a slice of the Higgs  branch (or mixed Coulomb-Higgs branch)
corresponding to a single instanton centered at the origin, $y^m
=0$.

In ``singular gauge'', the $SU(2)$ instanton (using the same
conventions as \cite{Dorey:2002ik}) is given by
\begin{align}\label{thinst}
A_\mu = 0,\qquad A_m = \frac{2Q^2 \bar\sigma_{nm}y_n}{y^2(y^2 +
Q^2)},
\end{align}
\newpage
\noindent
where $Q$ is the instanton size, and
\begin{align} \bar\sigma_{mn} &\equiv \frac{1}{4}(\bar\sigma_m \sigma_n -
\bar\sigma_n \sigma_m) \, , \qquad \sigma_{mn} \equiv
\frac{1}{4}(\sigma_m \bar\sigma_n - \sigma_n \bar\sigma_m), \nonumber \\
 \sigma_m &\equiv (i\vec\tau, 1_{2\times 2}), \qquad
\bar\sigma_m \equiv \sigma_m^{\dagger} = (-i\vec\tau, 1_{2\times
2})\,.
\end{align}
with $\vec\tau$ being the three Pauli-matrices. We choose singular
gauge, as opposed to the regular gauge in which $A_n = 2
\sigma_{mn}y^m/(y^2 + Q^2)$,  because of the improved asymptotic
behavior at large $y$. In the AdS setting,  the Higgs branch
should correspond to a normalizable deformation of the AdS
background at the origin of the moduli space. We see that the
solution falls to zero at large $y$ and the deformation in the
interior is controlled by the dimension two parameter $Q^2$ which
corresponds to the VEV of $\tilde q^i q_i$.  The choice of
singular gauge will be of particular use when computing the
effects of the chemical potential. The singularity of
\eqref{thinst} at $y^m =0$ is not problematic for computations of
physical (gauge invariant) quantities.

For $m=0$ and zero instanton number,  the background geometry
preserves the symmetries $SO(2,4)\times SU(2)_L \times SU(2)_R
\times U(1)_R \times SU(2)_f$. The $SO(2,4)$ isometry of the
$AdS^5$ factor corresponds to the conformal symmetry of the dual
gauge theory. The $SU(2)_L \times SU(2)_R \sim SO(4)$ acts in the
obvious way on the coordinates $y^m$. The $SU(2)_L$ factor
corresponds to a global symmetry, while $SU(2)_R$ and $U(1)_R$,
which acts on $(Z^1,Z^2)$, correspond to
R-symmetries\footnote{There is no apparent distinction between
$SU(2)_L$ and $SU(2)_R$ in the geometry of the AdS background. The
distinction arises from the gauge field couplings to the RR
four-form.}. $SU(2)_f$ is a gauge transformation on the D7-branes
which is constant at the boundary of AdS. The action at the
boundary of AdS corresponds to the flavor symmetry of the dual
gauge theory.

In the presence of the instanton \eqref{thinst}, and for $m\ne 0$,
the symmetries are broken to $SO(1,3) \times SU(2)_L \times {\rm
diag} (SU(2)_R \times SU(2)_f)$  corresponding to a point on the
mixed Coulomb-Higgs branch
\begin{align}\label{point}
q_{i\alpha} =
\begin{pmatrix} 0 \cr \vdots\cr 0\cr \epsilon_{i\alpha}Q
\end{pmatrix}\, , \,\,\, \phi_\alpha =
\begin{pmatrix}0 & & & &\cr & \ddots & &  &\cr & & & &\cr & & & & 0
\end{pmatrix}\, , \,\,\, \phi_3 =
\begin{pmatrix}0 & & \cr & \ddots &  & & \cr & & 0 & \cr & & & -m
\end{pmatrix},
\end{align} where $q_{i\alpha}$ are the scalar components of the
fundamental hypermultiplets,  with flavor index $i =1,2$ and
$SU(2)_R$ index $\alpha =1,2$. The scalars $\phi_\alpha$ belong to
the adjoint hypermultiplet, while $\phi_3$ is the adjoint scalar
belonging to the vector-multiplet.

In general, one expects that supersymmetry breaking deformations,
such as a finite chemical potential or temperature, lift the
vacuum degeneracy of the Higgs and Coulomb branch. In the dual
supergravity description,  the lifting of the Higgs branch at
finite temperature occurs because the conspiracy between the
Wess-Zumino and Yang-Mills terms, necessary for self-dual field
strengths to be solutions, no longer occurs. The potential which
is generated on the Higgs branch can be computed by evaluating the
D7-brane action on the space of self-dual field strengths.  In the
following we will compute the potential $V(Q)$ generated on the
slice of the Higgs branch dual to the single instanton
\eqref{thinst}.

\section{Finite chemical potential, zero temperature}

As the simplest example of a potential generated on the Higgs
branch, we first consider the case of  finite chemical
potential and zero temperature. For two flavors, there is a
U(2) flavor symmetry, with Cartan
generators $1$ and $\sigma_3$, corresponding to baryon number and
isospin respectively.  We will consider a nonzero chemical
potential for the isospin\footnote{Turning on a chemical potential
for the baryon number has no apparent effect here since none of the fields
on the D7 brane world volume are charged under this symmetry. This was
pointed out to us by A.~Karch.}.
To include the chemical potential \cite{Harnik:2003ke}
we allow a spurious gauge
field associated with the $\tau^3$ component of isospin
to acquire a VEV, $\mu$, in its $A^0$ component. This includes generic
fermion and scalar Lagrangian terms for fields with isospin charge
$e$ of the form
\begin{equation} \delta {\cal L} = - \mu e \bar{\psi} \tau^3 \gamma^0 \psi
+ \mu^2 e^2 |\phi|^2 \, .\end{equation}
The first term is a source for the fermionic isospin number density.  In
the path integral, this term places  the theory at finite density. The second term
is an unbounded scalar potential which renders the theory
unstable, such that Bose-Einstein condensation is expected. We will reproduce this run-away
behaviour of the scalar potential using the AdS/CFT description.
Equivalently, we may set $A_\tau = 0$ by a
gauge transformation $e^{i\mu e \tau}$, and perform the path
integral with boundary conditions corresponding to a spinning
configuration space.

AdS/CFT duality relates global symmetries of the boundary theory
to gauge symmetries in the bulk supergravity. The prescription for
turning on a chemical potential in the AdS description is to turn
on a background (non-normalizable) flat gauge connection ${\cal
A}_\tau = \mu$ for the associated gauge symmetry.  Equivalently,
one may consider a spinning AdS background. Examples of the AdS
description of a chemical potential in various contexts have
appeared in
\cite{Csaki:1998cb}-\cite{Evans:2001ab}.

The effect of the chemical potential on the Higgs branch can be
studied in the AdS description by computing the action for rigidly
rotating instantons with moduli ${\cal M}^i$,
\begin{align}\label{rigidone}
A_n = e^{i\mu t \sigma_3} {\cal A}_n^{\rm instanton}(y^m, {\cal
M}^i) e^{-i\mu t \sigma_3} \, .
\end{align}
Equivalently, we may add a background $A^0$ in the fixed
instanton background,
\begin{align}\label{backg}A^0 =
\begin{pmatrix} \mu & 0 \cr 0 & -\mu \end{pmatrix}, \qquad A_n =
A_n^{\rm instanton} \, .
\end{align}
On the slice of the Higgs branch corresponding to the single
instanton configurations \eqref{thinst} with modulus $Q$, the
effective potential
at quadratic order in $\mu$ can be determined by inserting
\eqref{backg}
\newpage
\noindent
into the D7-brane action $\int d^4 x \; V(Q) = -S_{D7}$
giving
\begin{eqnarray}\label{strx} V(Q) =
 T_7 \frac{(2\pi \alpha')^2}{g_s}
\int \, d^4 y\, {\rm tr}\left( \frac{1}{2}\frac{(y^2 +
m^2)^2}{ R^4} F^-_{mn}F^-_{mn} + 2F_{m \mu}F_{m \nu}\eta^{\mu\nu}
\right.\\ \left.
+
\frac{R^4}{(y^2 +
m^2)^2}F_{\mu\nu}F_{\alpha\beta}\eta^{\mu\alpha}\eta^{\nu\beta}\right)\,
, \nonumber
\end{eqnarray} with $y^2 = y^m y^m$. We have split the action into the
pieces involving $F$ in the $x$ and $y$ directions, indicated by
Greek and Roman indices respectively, as well as mixed terms. For
the background \eqref{backg},  the only non-zero contribution to
the potential comes from the mixed term ${\rm tr}\, F_{\mu
m}F_{\nu m}\eta^{\mu\nu} = -{\rm tr}\, [A_0, A_n]^2$, giving
\begin{align}\label{chempot}
V(Q) = -
T_7 \frac{2(4\pi\alpha')^2}{g_s}\mu^2\int d^4y \frac{
Q^4}{y^2(y^2+Q^2)^2} = -
T_7 \frac{2(4\pi^2 \alpha')^2}{g_s}
\mu^2Q^2 \, .
\end{align}

Note that the term $\sqrt{g}F_{\mu m}F^{\mu m}$ in the D7-action
is also the term which determines the metric (two-derivative
effective action) on the Higgs branch. To quadratic order in the
chemical potential, the result (\ref{chempot}) is exact.  Higher
dimension operators in the DBI action with two Greek indices,
which have been omitted from (\ref{strx}), must vanish in the
instanton background, otherwise there would be corrections to the
Higgs branch metric in a strong 't Hooft coupling expansion
\cite{Guralnik:2004ve,Guralnik:2004wq}. By the same token, the
term in the potential on the Higgs branch which is quadratic in
the chemical potential is given exactly by the $F_{\mu m}F^{\mu
m}$ term.

In light of (\ref{chempot}),  the ${\cal N}=2$ theory is unstable
at finite chemical potential. Physically, the instability is due
to Bose-Einstein condensation. This instability will not be cured
by taking into account the higher derivative terms in the
effective action, whose contribution does not effect the large $Q$
asymptotics. Potentially, the instability could be removed by the inclusion of
an appropriately large mass term for the scalars, either as an
explicit additional perturbation, or generated radiatively
from the inclusion of some other relevant perturbation.
The finite temperature configurations studied below, however, will not
be able to cure this instability since - as shown below -
the relevant deformation, $T^4$, has dimension four
and can at best generate a potential $T^8/Q^4$. This does not stabilize the theory.

\section{Finite temperature, zero chemical potential}

Before investigating the large 't Hooft coupling behavior of the
effective potential on the Higgs branch,  let us review the
situation to one loop in perturbation theory.  In a supersymmetric
theory,  the one-loop effective potential is
\begin{align}
V= V_0+ \frac{1}{8}T^2 \sum_{i,j}\left|\frac{ \partial^2 W
}{\partial \phi_i
\partial\phi_j}\right|^2 \, ,
\end{align}
where $V_0$ is the zero temperature effective potential, $W$ is
the superpotential, and $\phi_i$ are the scalar components of
chiral superfields.  For the theory we are considering,  this
potential drives the system towards the origin of moduli space.  In fact
it is often said that supersymmetric theories with flat directions
do not have symmetry breaking vacua at finite temperature
\cite{Haber:1982nb} (for some exceptions, see
\cite{Dvali:1996fz,Dvali:1998ct}). However, for large 't Hooft
coupling,  we will find that the theory we consider does indeed exhibit
symmetry breaking at high temperature.  The origin of the Higgs
branch becomes a maximum of the effective potential for $T>T_c$.

The gravitational dual of ${\cal N}=4$ gauge theory at large 't
Hooft coupling and finite temperature is obtained by replacing the
background \eqref{adsgeo} with the AdS-Schwarzschild black-hole
background \cite{Witten:1998zw}. The latter belongs to a general
class of supergravity solutions which, in a choice of coordinates
convenient for our purposes, have the form
 \begin{align}
 \label{bhgeneral}
ds^2 &= f(r)(d\vec x^2 + g(r)d\tau^2)+
h(r)(\sum_{m=1}^4 dy^m dy^m + \sum_{i=1}^2 dZ^i dZ^i),\nonumber\\
e^{-\Phi}&=\phi(r), \nonumber \\[2.5mm]
F^{(5)} &= 4 R^4 (V_{S^5} + ^*V_{S^5})=dC_{(4)}, \qquad
C_{(4)}|_{0123} = s(r)\, dx^0\wedge
dx^1\wedge dx^2\wedge dx^3 \, ,   \nonumber\\[3mm]
r^2 &= y^m y^m + Z^i Z^i.
\end{align}
For the AdS-Schwarzschild solution, we have \begin{align}\label{AdSSch} f(r)
= \frac{4r^4 + b^4}{4r^2R^2},\,\,\,\,  g(r) =
\left(\frac{4r^4-b^4}{4r^4+b^4}\right)^2,\,\,\,\, 
 h(r) =
\frac{R^2}{r^2},\,\,\,\, s(r) = \frac{r^4}{R^4}\left(1+
\frac{b^8}{16r^8}\right)\, . \end{align} The coordinates $\vec x$ are
the spatial coordinates of the dual gauge theory and $\tau$ is the
Euclidean time direction, which is compactified on a circle of
radius $b^{-1}$, corresponding to the inverse temperature. Note
that the temperature $T\sim b$ only enters to the fourth power.

The D7 embedding in this background to describe our ${\cal
N}=2$ gauge theory with quarks is
\begin{align}\label{gino}
Z^2=0 \, ,\qquad Z^1 = z(y),\end{align} where $ y^2 \equiv
 y^my^m.$  The induced metric $G|_{\rm pb}$ takes the form
\begin{align}ds^2_{\rm D7} = f(r)\left(d\vec x^2 +
g(r)d\tau^2\right)+ h(r)\left(dy^mdy^m+
\left(\frac{z'(y)}{y}\right)^2(y^m dy^m)^2\right),\end{align} with
$r^2 = y^2 + z^2(y)$.  To lowest order in $\alpha'$, the D7-brane
action is
\begin{equation}
 -T_7 \int d^8 \xi \; e^{-\Phi}\sqrt{-\det G}|_{\rm pb}
 =-2\pi^2 T_7\int d^4 x \; dy \; y^3 \phi(r) f(r)^2 \sqrt{g(r)}\; h(r)^2 \sqrt{1+z'(y)^2} , \label{try}
\end{equation}
and the specific embeddings $z(y)$ can be found by solving the
Euler-Lagrange equation
with boundary conditions $z(\infty)=m,$ $z'(\infty)=0.$


At large $y$ the geometry returns to AdS
where the asymptotic embedding solution (\ref{gino}) takes the form  $z(y) \sim
m + c/y^2$. Given $m$, the parameter $c$, corresponding to a
$\bar\psi\psi$ condensate, is determined by requiring the solution
to be smooth and normalizable.  These solutions were studied in
detail in \cite{BEEGK}.

The potential generated on the Higgs branch is obtained by
calculating a contribution to the D7-brane action involving the
gauge field $F_{ab}$
 at second order in $\alpha'$.
Specifically, one evaluates the action on the space of field
strengths which are self-dual\footnote{There are couplings between
world-volume scalars and field strengths at higher orders in
$\alpha'$ which could alter the embedding. However we only
consider the leading term in a large 't Hooft coupling expansion
for which these couplings can be neglected. 
} with respect to the induced metric in the directions transverse
to $\tau,\vec x$;
\begin{align}\label{thepot}
V = {T_7 ( 2 \pi \alpha')^2 \over 2}\left( \frac{1}{g_s} \int d^4y\,
C^{(4)}_{0123}\, \epsilon_{mnrs}\,{\rm Tr}\, F_{mn}F_{rs}
-\frac{1}{2} \int\, d^4y\, e^{-\Phi}\sqrt{-{\rm det}G}\,{\rm
Tr}\,F^{mn}F_{mn}\right),
\end{align}
where 
$F_{mn}$ is self-dual with
respect to the metric
\begin{align}\label{sdualmet}ds_\perp^2 = h(r)\left((1+z'(y)^2) dy^2+
y^2 d\Omega_3^2\right)\, .\end{align} This metric is conformally
flat. With new coordinates $\tilde y(y)$ such that $ds^2 =
\alpha(\tilde
 y)(d\tilde y^2 + \tilde y^2d\Omega_3^2)$, the instanton
configurations (self-dual field strengths) take the usual form.

In the $b=0$ (zero temperature) limit,  the analysis of section
(\ref{sughiggs}) holds and the potential \eqref{thepot} becomes
flat. At finite temperature,  the conspiracy between the
Wess-Zumino term and the Yang-Mills terms fails, giving a
potential on the moduli space of instantons. We now compute the
form of the potential for the slice of moduli space corresponding
to an instanton centered at the origin.

\subsection{Computation of the Higgs Effective Potential}

We wish to evaluate the D7-brane action on the space of fields
strengths which are self-dual with respect to the transverse part
of the metric $$ds_\perp^2 = h(r)\left( dy^mdy^m+
\left(\frac{z'(y)}{y}\right)^2(y^m dy^m)^2\right).$$ This metric
is conformally flat. There exists a coordinate transformation
\begin{equation}\tilde y^m =J(y)y^m \end{equation} such that the
induced metric has the form
\begin{align}ds^2_{\rm D7} &=
f(r)\left(d\vec x^2 + g(r)d\tau^2\right)+
\frac{h(r)}{J(y)^2}d\tilde y^md\tilde y^m\, .\end{align} The
function $J(y)$ is obtained by solving
\begin{align}\label{solvethis}\left(\frac{J'}{J}\right)^2 + \frac{2}{y}\frac{J'}{J}
 - \frac{1}{y^2}z'(y)^2 = 0  \, , \end{align}
subject to the boundary condition $J(y\rightarrow\infty) = 1$. In
the coordinates $\tilde y^m$, the single instanton centered at the
origin has the usual form,
\begin{align}
A_m = \frac{2Q^2\bar\sigma_{nm}\tilde y^n}{\tilde y^2(\tilde y^2 +
Q^2)}\, . \end{align} The embedding function $z(y)$ is determined
by shooting techniques as in \cite{BEEGK}. The
potential on the Higgs branch is then given by $S_{D7} = - \int d^4
x V(Q)$, or
\begin{align}\label{prepot}V(Q) &=
T_7 {( 2 \pi \alpha')^2 \over 2}\left(\frac{1}{g_s} \int d^4y\,
C^{(4)}_{0123}\, \epsilon_{mnrs}\,{\rm Tr}\, F_{mn}F_{rs}
-\frac{1}{2} \int\, d^4y\, e^{-\Phi}\sqrt{-{\rm det}G}\,{\rm
Tr}F^{mn}F_{mn}\right)  \nonumber \\[2mm]
&= T_7 {( 2 \pi \alpha')^2 \over 4 g_s} \int d^4 \tilde y \left( s(r) - \phi(r) f(r)^2\sqrt{g(r)}\right)
\frac{96\, Q^4}{(\tilde y^2 + Q^2)^4} \\[2mm]
&=T_7 {( 2 \pi \alpha')^2 \over 4 g_s}  Vol(S_3)\int y^3 dy  \left( s(r) - \phi(r) f(r)^2\sqrt{g(r)}\right)
\frac{96 \,Q^4 \, J(y)^4}{(J(y)^2y^2 + Q^2)^4} \left(1+y
\frac{J'(y)}{J(y)}\right).\nonumber
\end{align}
For the AdS-Schwarzschild background
(\ref{bhgeneral},\ref{AdSSch}) we obtain
\begin{equation}
\label{bhpotentialinr}
V(Q)\,=\, T_7 {6 ( 2 \pi^2 \alpha')^2 \over g_s R^4 }  \int dy \, { b^8 y^3 \over (J(y)^2 y^2+z(y)^2)^2}
\frac{Q^4 J(y)^4}{(J(y)^2y^2 + Q^2)^4} \left(1+y
\frac{J'(y)}{J(y)}\right).
\end{equation}

\subsection{Large VEV Asymptotics}

Without resorting to numerics, some qualitative features of $V(Q)$
at large $Q$ (or large instanton size) can be determined from the
large $y$ behavior of the induced metric and RR-form given in
(\ref{bhgeneral},\ref{AdSSch}), together with the asymptotic
behaviour of the embedding,
\begin{equation}\label{asembed} z(y) = m + \frac{c}{y^2} + \cdots \, .
\end{equation}  At large $y$ we can neglect the condensate and the leading
terms in the potential will depend on $m$ and on the deformation parameter of
the background $b$. The D7-brane
action for a large instanton, which has support at large $y$ is
\begin{align} \label{bim} \int d^4 x V(Q) &= - S_{D7} \nonumber \\[1mm]
 &\approx T_7
\frac{6(2\pi^2\alpha')^2}{g_s R^4} \int d^4x \int dy\;
y^3 {Q^4 \over (y^2 + Q^2)^4} \left[ {b^8 \over  y^4} -
{2 b^8 m^2 \over   y^6} +...\right]\\[3mm] &
= T_7 \frac{6(2\pi^2\alpha')^2}{g_s R^4} \left[- {b^8 \over
Q^4} \left(\frac{1}{2}\log{2}-\frac{1}{3}\right) + {2 m^2 b^8 \over
Q^6} \left( \frac{67}{48}-2 \log{2}\right) + \ldots  \right].
\nonumber\end{align}
This
asymptotic approximation is valid when $y \ge Q$ but gives IR
divergent integrals. We have evaluated the integrals down to
$y/Q=1$. In the next section we evaluate the full potential
including the correct, convergent, IR behaviour.

The asymptotic potential terms are simply what one would expect on
dimensional grounds. The potential must vanish as $b\rightarrow 0$
and the geometry becomes AdS. Since $b$ enters as $b^4$, the first
Q dependent term must take the form $b^8/Q^4$.

\subsection{Numerical computation of $V(Q)$}

\begin{figure}[H]
\begin{center}
\includegraphics[scale=.8]{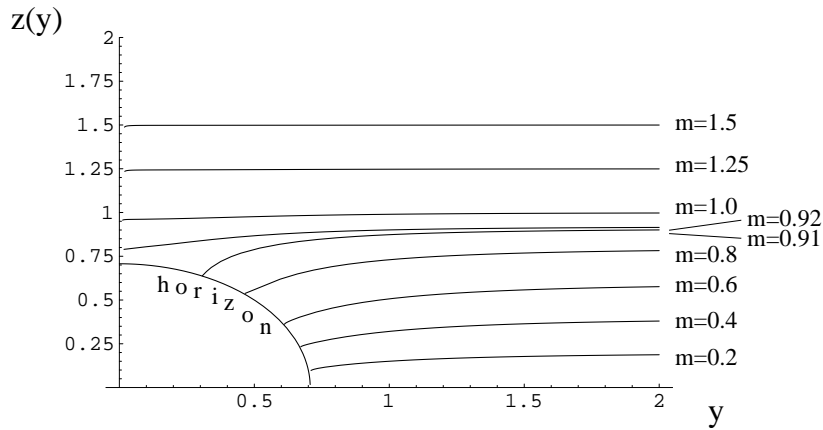}
\caption{Brane embeddings in the AdS Schwarzschild background for different
  values of the quark mass. In the plot we have set $b=1$.}
\label{d7inbh}
\end{center}
\end{figure}

To compute $V(Q)$ in general requires knowledge of the embedding
function $z(y)$,  which has been computed by a numerical
shooting technique in \cite{BEEGK}.  Imposing boundary conditions
corresponding to the large $y$ behaviour \eqref{asembed}, and
requiring smooth behaviour in the interior, such that an RG flow interpretation is possible, leads to a dependence $c(m,b)$ of the chiral quark
condensate $\langle\bar\psi\psi\rangle$ on the quark mass and on the temperature.
Depending on the ratio $m/b$,  there are two types of solutions,
which differ  by the topology of the D7-branes.  At large $r$ (or
$y$) the geometry of the D7-branes is $AdS_5 \times S^3$ and
the topology of the $r\rightarrow\infty$ boundary is $S^1 \times
R^3 \times S^3$. For sufficiently large $m/b$, the $S^3$ component
of the D7-geometry contracts to zero size at finite $r > b$. In
this case the D7-brane ``ends'' before reaching the horizon at
$r=b$. However, for sufficiently small $m/b$,  the D7-brane ends
at the horizon, at which point the thermal $S^1$ contracts to zero
size.
Both these types of solutions are plotted in figure \ref{d7inbh}.
There is a first order phase transition at the critical value of
$m/b \approx 0.92$ where the two types of solution meet
\cite{BEEGK,Kirsch:2004km}.  The $\langle\bar\psi\psi\rangle$ condensate is
non-zero on both sides of this transition, although there is a
discontinuous jump in its value.
\\
%
%

Using the machinery discussed above,  we can  study the behaviour
of the scalar VEV parametrizing the Higgs branch on either side of
the transition.  We evaluate the potential
(\ref{thepot},\ref{bhpotentialinr}) numerically, with a single
instanton centred at the origin  for different values of the ratio
$m/b$. The results are plotted in figure \ref{potinbh}.
\setlength{\unitlength}{1mm}
\begin{center}
\begin{figure}[H]
\begin{picture}(130,70)(0,0)
\put(0,3){\epsfig{figure=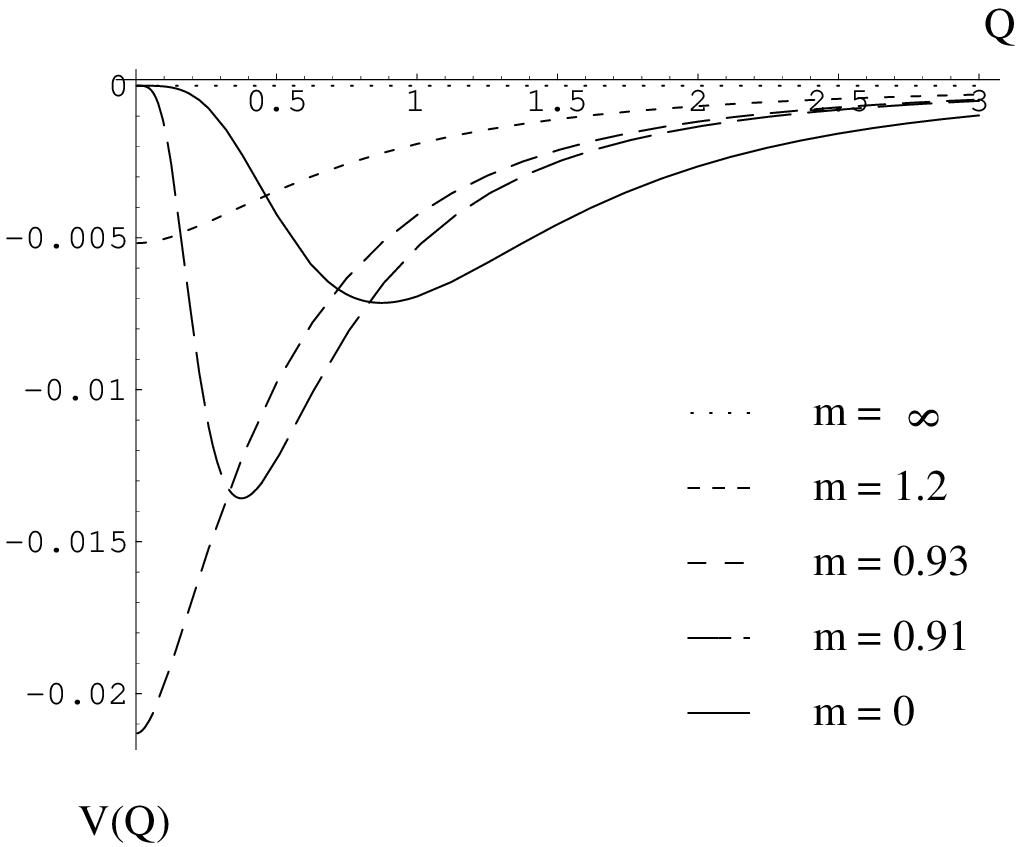,width=7.5cm}}
\put(90,3){\epsfig{figure=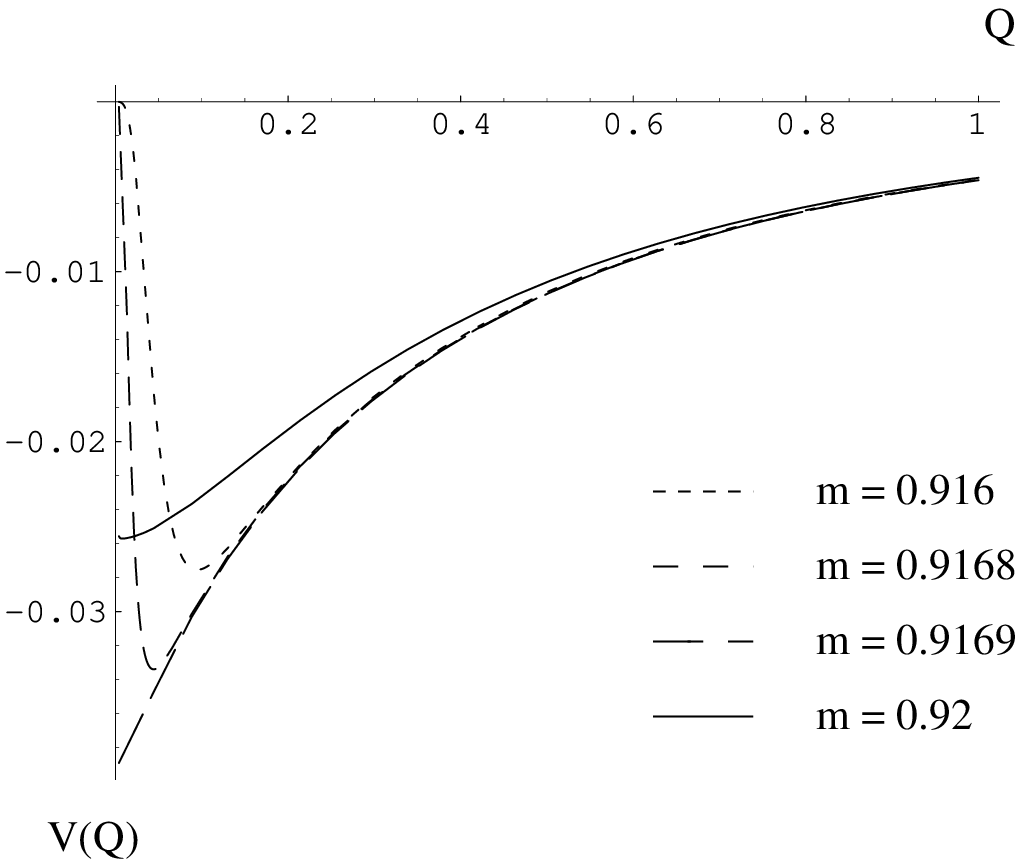,width=7.5cm}}
\end{picture}
\caption{Potential $V(Q)$ as a function of the instanton size /
Higgs VEV $Q$ for various values of the quark mass $m$ (we set
$b=1$ here); in fig. a)  we sample the whole spectrum from $m=0$
to $m\to \infty$, in fig. b) we vary $m$ close to the phase
transition region. The potential for  $m \to \infty$ is flat and coincides
with the horizontal axis.} \label{potinbh}
\end{figure}
\end{center}

\begin{center}
\begin{figure}[H]
\begin{picture}(200,56)(0,0)
\put(5,5){\epsfig{figure=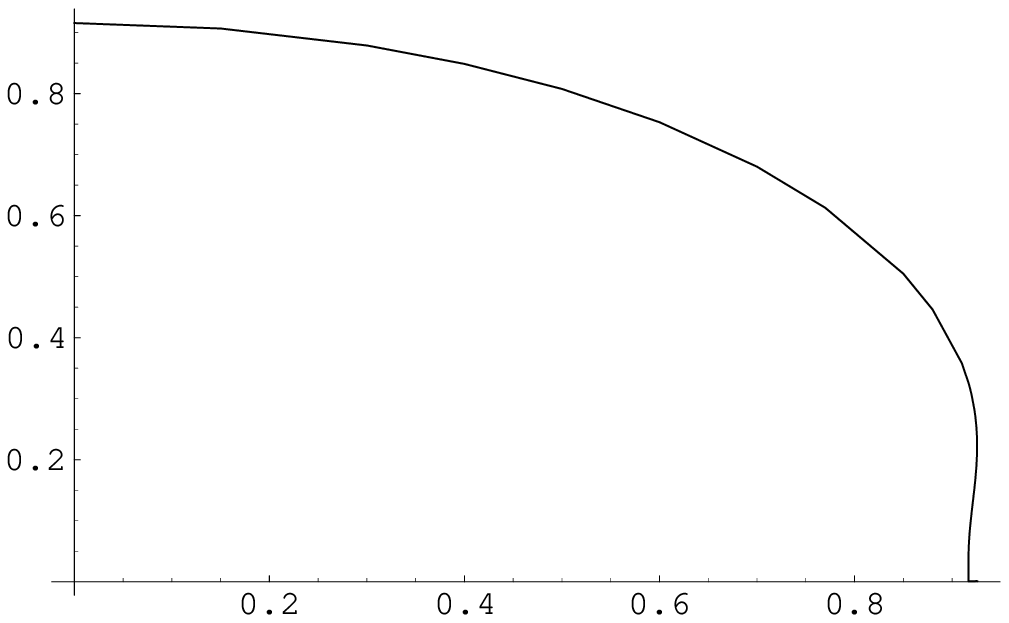,width=8cm}}
\put(103,5){\epsfig{figure=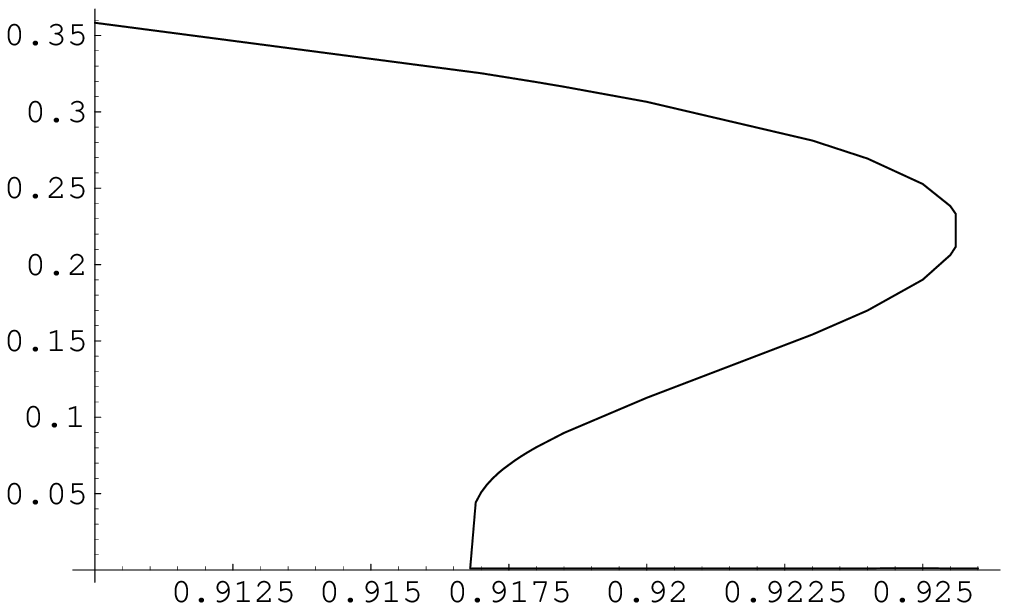,width=7cm}} \thinlines
\put(78,7){\line(1,0){7}} \put(78,7){\line(0,1){21}}
\put(85,28){\line(-1,0){7}} \put(85,28){\line(0,-1){21}}
\put(80,0){\makebox(0,0)[b]{$m_q/T$}}
\put(2,52){\makebox(0,0)[b]{$Q_{0}$}}
\end{picture}
\caption{Position of the minimum of the potential $Q_0$ versus the
bare quark mass $m_q$, zoom of
  the critical region. }
\label{vevvsm}
\end{figure}
\end{center}

Figure \ref{potinbh} shows that there are two phases for the
scalar Higgs VEV matching the two phases (topologies) of the
D7-brane embedding.  These phases are divided by the first order
phase transition at a critical value of $m/b$, where the parameter
$b$ of the AdS/Schwarzchild  background is proportional to the
temperature. In figure \ref{potinbh} we have chosen units such
that $b=1$. For $m/b \to \infty$, the D7 branes probe the AdS-like
part of the geometry, such that the potential is flat (dotted line
on the left hand side of figure \ref{potinbh}). For all other
values of $m/b$, the potential approaches this constant asymptotic
value from below for large $Q$. When the value of $m/b$ is
decreased, starting at $m/b\to \infty$, a minimum of the potential
forms at $Q=0$, with increasing depth.  At the critical value
$m_c/b \simeq 0.92$, a new solution for the brane embedding
appears, for which the potential has a minimum at $Q \neq 0$. For
$m/b \to 0$, $V$ has a maximum at $Q =0$ and a minimum of
decreasing depth for $Q \neq 0$. At $m/b=0$ (solid line in the
plot) there is still a shallow minimum of $V$ at $Q \simeq 0.85
b$. Thus, the Higgs VEV is an order parameter for the first order
phase transition.  We emphasize that the symmetry breaking phase
occurs at small $m/b$, or high temperature, contrary to the usual
expectation.   We plot the value of the Higgs VEV  as a function
of $m/b$ in Figure \ref{vevvsm}.
\vspace*{-1cm}

Note that the expectation value $Q_0$ is a multivalued function of
$m/b$ in a region near the first order critical point. There are
several regular embeddings in this region,  with the true vacuum
corresponding to the solution with lower free energy.


Thus we find that for $T\geq T_c$, the theory is driven to a Higgs
phase or, more precisely, a point on the mixed Coulomb-Higgs
branch.  We have not determined whether or not this vacuum is
merely metastable, which would be the case if the origin of the
Coulomb branch has lower free energy.

Since the less supersymmetric phase occurs for $T \ge T_c$,
the phase transition discussed here is quite different from the
deconfinement transition. Nevertheless it is interesting to note
that our phase transition is first order, as is found in lattice
simulations of the deconfinement transition in large $N$ $SU(N)$
gauge theories \cite{teper1}.


\subsection{Geometric aspects of the transition}

\begin{figure}[h]
\begin{center}
\includegraphics[scale=.78]{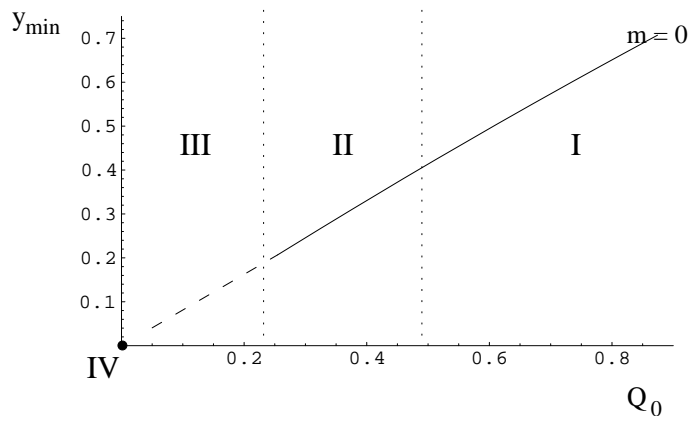}
\caption{
Position of the
endpoint $y_{min}$ of the embedded brane  on the horizon versus the potential minimum $Q_0$. 
Region I corresponds to solutions with $0\le$ $m_q$ $< 0.91$. In this range, for each $m_q$ there is only one regular D7 embedding, reaching the horizon.
In the phase transition region ($0.913 \lesssim m_q \lesssim 0.926$), there are three embeddings for each $m_q$ (see the text for details). Of the two that reach the horizon one lies on the straight line in region II, the other in region III (dashed line).${}^8$
The solution that does not touch the horizon has $y_{min}=Q_0=0$, thus lying at the origin IV (we use a dot to describe this kind of solutions).
Finally for large enough quark masses ($m_q>0.93$), there is again only one regular embedding for each $m_q$, which does not reach the horizon.
These solutions all accumulate in the origin IV.
}
\label{vevvsrmin}
\end{center}
\end{figure}

It is worth stressing the link between the transition of the Higgs
VEV between the two phases and the change in the embedding
topology of the D7 brane.  From figures 1 and 2, we see that the
Higgs potential only has a non-zero minimum when the D7-branes
reach the horizon.  There is a repulsive effect in the vicinity of
the horizon which causes the instanton to expand.  In figure
\ref{vevvsrmin} we plot the value of the Higgs VEV versus the
minimum value of the coordinate $y$ to which the D7 branes extend.
When the D7-branes do not reach the horizon, the $S^3$ component
of their geometry contracts to zero size at the endpoint, which
occurs at $y=0$. On the other hand, when the D7-branes reach the
horizon, the $S^3$ component of their geometry remains finite size
and $y\ne 0$ at the endpoint. The dependence of the VEV on the
minimum value of $y$ is almost a straight line with slope roughly
equal to 1.25. The solutions that reach the horizon are
distributed along this line, while the ones which end before
reaching the horizon accumulate at the origin.

\begin{center}
\begin{figure}[h]
\begin{picture}(200,40)(0,0)
\put(1,-1){\epsfig{figure=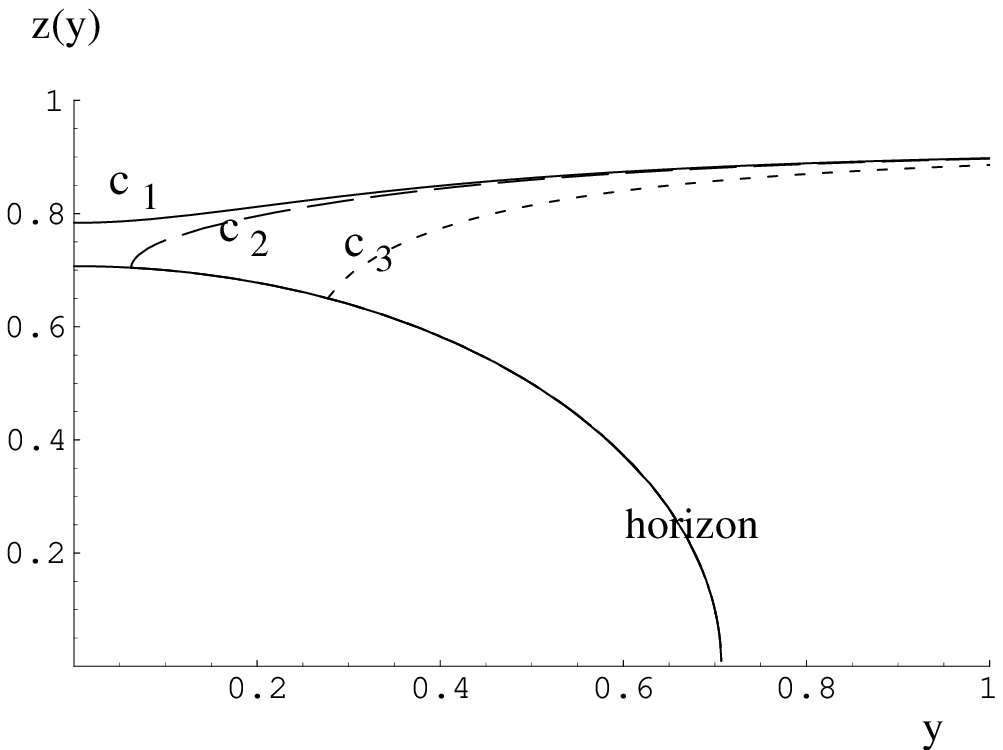,width=7.8cm}}
\put(88,3){\epsfig{figure=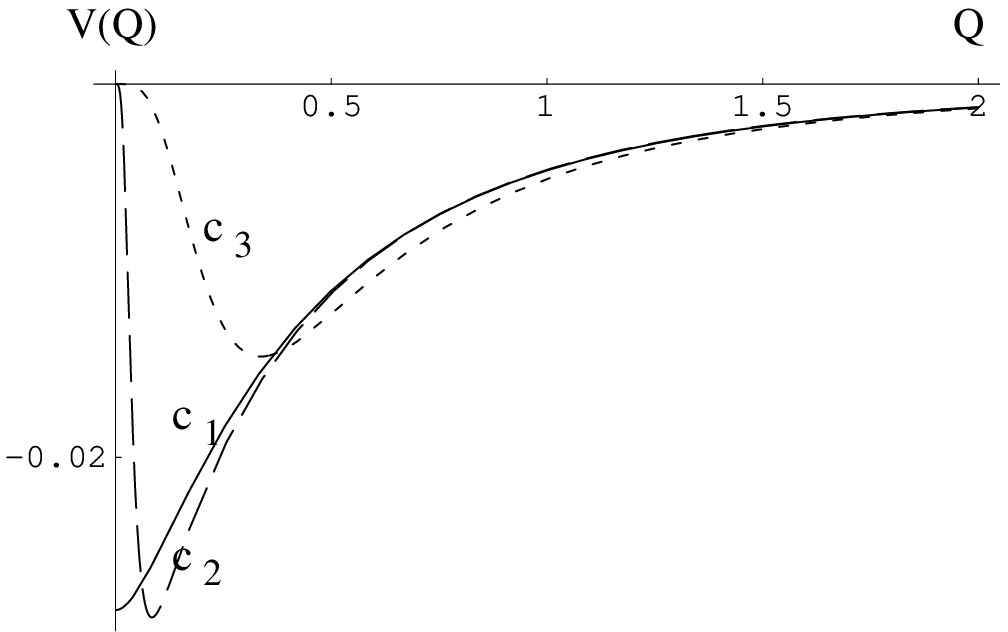,width=8.5cm}}
\end{picture}
\caption{Solutions in critical region: There are three condensate
values which
  give different regular
  solutions for the each quark mass. $m_q=0.918$ in the plot with condensate
  values $c_1=-0.0165$, $c_2=-0.0172$, $c_3=-0.0265$. }
\label{samemdiffc}
\end{figure}
\end{center}

The phase transition region ($m/b\simeq 0.91$) has some additional
interesting structure. In~this region there are three regular
solutions for the D7 embedding for each value of $m_q$ (see
\cite{Kirsch:2004km,Kruczenski:2003uq}). Some of them end on the
horizon and some do not. The three flows are shown for a
particular case in fig. \ref{samemdiffc}a. These different
embeddings give different predictions for the Higgs VEV. We~also
show the potential for $Q$ on each of these embeddings in figure
\ref{samemdiffc}b and the minimum vs $m$ can~be found in figure
\ref{vevvsm}b. There are three values of $Q$ supplied in this
transition region. Clearly there will be a first order transition
where the embedding jumps from solutions ending on and off the
horizon as $m$ is changed, corresponding to a discontinuous
transition in the value of the scalar VEV.

\section{Conclusions}

We have investigated the thermodynamics of a large $N_c$ strongly
coupled ${\cal N}=2$ gauge theory, using AdS/CFT duality to
compute the potential which is generated on the Higgs branch at
finite temperature and isospin chemical potential.  In the AdS
description, this potential is a potential on the moduli space of
instantons.

We have shown non-perturbatively that the chemical potential
destabilizes the Higgs potential, giving rise to Bose-Einstein
condensation.  The ${\cal O}(\mu^2)$ term in the potential can be
computed exactly,  as a consequence of the non-renormalization of
the metric on the Higgs branch,  which is correctly reproduced by
instanton dynamics in the AdS description. The instability due to
the chemical potential is not cured at finite temperature.

It would be very interesting to introduce a chemical potential for
Baryon number in the AdS setting. Unlike the chemical potential
for isospin,  this is known to be very difficult to study on the
lattice due to a complex fermionic determinant.  In the AdS
setting,  it is also easier to introduce a chemical potential for
the isospin, as we have done here, although the problems arising
for baryon number are probably much more tractable than those
which arise on the lattice.

\newpage

\vspace{2.cm}
\footnotetext[8]{There are two small gaps in the plot, one between region II and III and one between III and IV. It is worth stressing that quark masses for solutions in these areas all lie in two very narrow ranges around $m_q \simeq 0.925$ and $m_q \simeq 0.916$.  At a certain point the fine-tuning required to find the solutions which should fill the gaps  exceeds the numerical accuracy used to solve the equation of motion.}
%
\newpage

It was known previously that this theory has a first order phase
transition at a critical temperature of order of the quark mass,
and that the chiral quark condensate $\langle\bar\psi\psi\rangle$
varies discontinuously across the transition
\cite{BEEGK,Kirsch:2004km,Kruczenski:2003uq}. By computing the
effective potential on a mixed Coulomb-Higgs branch, we have found
that the  Higgs VEV is an order parameter for this transition. The
VEV is zero for $0<T\le T_c$ and non-zero for $T>T_c$. For
vanishing quark mass, in which case a pure Higgs branch with
vanishing Coulomb-branch moduli exists, the effective potential
suggests that the finite temperature theory is in a (stable) Higgs
phase.  Although there are numerous examples of high temperature
symmetry breaking in non-supersymmetric theories,  our strong
coupling result differs from the typical weak coupling behavior of
renormalizable supersymmetric theories with flat directions, for
which the origin of moduli space becomes a stable or metastable
vacuum at finite temperature.

Another application of the machinery we have developed here would
be to compute the effective potential generated on the Higgs
branch by other supersymmetry breaking deformations of the ${\cal
N}=2$ theory, besides the thermal deformations we discussed here.
For example,  there has been considerable interest in deformations
that give rise to electrically confining theories. In principle,
for all such theories one should check that the scalar potential
has a minimum at the origin of the Higgs branch, before computing
physical observables such as meson spectra. Moreover, in the
future it might be possible with these techniques to reproduce the
Affleck-Dine-Seiberg runaway scalar potential of ${\cal N}=1$
supersymmetric QCD \cite{Davis:1983mz,Affleck:1983mk}, which
arises when the number of flavors is less than the number of
colors. \vspace{3cm}

\section*{Acknowledgements}

We thank G.~Burgio, A.~Cohen, A.~Karch and M.~Schmaltz for useful
conversations.  Z.G. is grateful to the Boston University visitor
program for hospitality during the completion of this work.

The work of J.E.~and Z.G.~has been funded by DFG (Deutsche
Forschungsgemeinschaft) within the Emmy Noether programme, grant
ER301/1-4. The work of R.~A. has been funded by DFG within the
`Schwerpunktprogramm Stringtheorie', grant ER301/2-1.

\end{document}